\documentclass[seceq,supplement]{ptptex}

\usepackage{graphicx}
\usepackage{subfig}

\title{
Elastic Scattering and Total Cross-Section in p+p reactions 
}

\subtitle{
as Measured by the LHC Experiment TOTEM  
at $\sqrt{s} = 7$ TeV 
}    

\author{
Tam\'as \textsc{Cs\"org\H{o}}$^{4}$, for the TOTEM Collaboration:\\
G.~Antchev\footnote{INRNE-BAS, Inst. for Nucl. Res. and Nucl. Energy, Bulgarian Acad. Sci., Sofia, Bulgaria}\addtocounter{footnote}{-1}, 
P.~Aspell$^{8}$, 
I.~Atanassov$^{8}$\footnotemark, 
V.~Avati$^{8}$, 
J.~Baechler$^{8}$,
V.~Berardi$^{5b,5a}$, 
M.~Berretti$^{7b}$, 
E.~Bossini$^{7b}$, 
M.~Bozzo$^{6b,6a}$,
P.~Brogi$^{7b}$, 
E.~Br\"{u}cken$^{3a,3b}$, 
A.~Buzzo$^{6a}$, 
F.~S.~Cafagna$^{5a}$, 
M.~Calicchio$^{5b,5a}$,
M.~G.~Catanesi$^{5a}$, 
C.~Covault$^{9}$, 
M.~Csan\'{a}d$^{4}$\footnote{Department of Atomic Physics, ELTE University, Budapest, Hungary} , 
M.~Deile$^{8}$, 
E.~Dimovasili$^{8}$, 
M.~Doubek$^{1b}$,
K.~Eggert$^{9}$, 
V.Eremin\footnote{Ioffe Physical - Technical Institute of Russian Academy of Sciences}, 
R.~Ferretti$^{6a,6b}$, 
F.~Ferro$^{6a}$, 
A. Fiergolski\footnote{Warsaw University of Technology, Warsaw, Poland}, 
F.~Garcia$^{3a}$, 
S.~Giani$^{8}$,
V.~Greco$^{7b,8}$, 
L.~Grzanka$^{8,}$\footnote{Institute of Nuclear Physics, Polish Academy of Science, Cracow, Poland}
\addtocounter{footnote}{-1}\addtocounter{footnote}{-1}  , 
J.~Heino$^{3a}$, 
T.~Hilden$^{3a,3b}$,
M.~R.~Intonti$^{5a}$, 
M.~Janda$^{1b}$, 
J.~Ka\v{s}par$^{1a,8}$, 
J.~Kopal$^{1a,8}$, 
V.~Kundr\'{a}t$^{1a}$, 
K.~Kurvinen$^{3a}$,
S.~Lami$^{7a}$, 
G.~Latino$^{7b}$,
R.~Lauhakangas$^{3a}$, 
T.~Leszko$^{\dagger}$,
E.~Lippmaa$^{2}$,
M.~Lokaj\'{\i}\v{c}ek$^{1a}$, 
M.~Lo~Vetere$^{6b,6a}$, 
F.~Lucas~Rodr\'{\i}guez$^{8}$,
M.~Macr\'{\i}$^{6a}$, 
L.~Magaletti$^{5b,5a}$,
G.~Magazz\`{u}$^{7a}$, 
A.~Mercadante$^{5b,5a}$, 
M.~Meucci$^{7b}$,
S.~Minutoli$^{6a}$, 
F.~Nemes$^{4,**}$,
H.~Niewiadomski$^{8}$, 
E.~Noschis$^{8}$, 
T.~Nov\'ak$^{4,}$\footnote{KRF, Gy\"{o}ngy\"{o}s, Hungary},
E.~Oliveri$^{7b}$, 
F.~Oljemark$^{3a,3b}$, 
R.~Orava$^{3a,3b}$, 
M.~Oriunno$^{8}$\footnote{SLAC National Accelerator Laboratory, Stanford CA, USA},
K.~\"{O}sterberg$^{3a,3b}$, 
P.~Palazzi$^{7b}$, 
A.-L.~Perrot$^{8}$, 
E.~Pedreschi$^{7a}$,
J.~Pet\"{a}j\"{a}j\"{a}rvi$^{3a}$, 
J.~Proch\'{a}zka$^{1a}$, 
M.~Quinto$^{5a}$,
E.~Radermacher$^{8}$, 
E.~Radicioni$^{5a}$,
F.~Ravotti$^{8}$, 
E.~Robutti$^{6a}$,
L.~Ropelewski$^{8}$, 
G.~Ruggiero$^{8}$,
H.~Saarikko$^{3a,3b}$,  
G.~Sanguinetti$^{7a}$,
A.~Santroni$^{6b,6a}$,
A.~Scribano$^{7b}$, 
G.~Sette$^{6b,6a}$, 
W.~Snoeys$^{8}$, 
F.~Spinella$^{7a}$,
J.~Sziklai$^{4}$, 
C.~Taylor$^{9}$,
N.~Turini$^{7b}$, 
V.~Vacek$^{1b}$, 
M.~V\'{i}tek$^{1b}$, 
J.~Welti$^{3a,b}$, 
J.~Whitmore$^{10}$.
}

\inst{
$^{1a}${Inst. of Physics, Academy of Sciences of Czech Republic, Praha, Czech Republic.}\\
$^{1b}${Czech Technical University, Praha, Czech Republic.}\\
$^{2}${National Institute of Chemical Physics and Biophysics NICPB, Tallinn, Estonia.}\\
$^{3a}${Helsinki Institute of Physics, Helsinki, Finland.}\\
$^{3b}${Department of Physics, University of Helsinki, Helsinki, Finland.}\\
$^{4}${MTA KFKI RMKI, Budapest, Hungary.}\\
$^{5a}${INFN Sezione di Bari, Bari, Italy.}\\
$^{5b}${Dipartimento Interateneo di Fisica di Bari, Bari, Italy.}\\
$^{6a}${Sezione INFN, Genova, Genova, Italy.}\\
$^{6b}${Universit\`{a} degli Studi di Genova, Genova, Italy.}\\
$^{7a}${INFN Sezione di Pisa, Pisa, Italy.}\\
$^{7b}${Universit\`{a} degli Studi di Siena and Gruppo Collegato INFN di Siena, Siena, Italy.}\\
$^{8}${CERN, Geneva, Switzerland.}\\
$^{9}${Dept. of Physics, Case Western Reserve University, Cleveland, OH, USA.}\\
$^{10}${Dept. of Physics, Penn State University, University Park, PA, USA.}\\
}

\abst{
 Proton-proton elastic scattering has been measured by the TOTEM
 experiment at the CERN Large Hadron Collider at $\sqrt{s} = 7 $ TeV  in special
 runs with  the Roman Pot detectors placed as close to the outgoing beam  as seven times the
 transverse beam size. 
 The differential cross-section measurements are reported  
 in the $|t|$-range of 0.36 to 2.5 GeV$^{2}$.  
 Extending the range of data to low $t$ values from 0.02 to 0.33 GeV$^2$,
 and utilizing the luminosity measurements of CMS,
 the total proton-proton cross section at $\sqrt{s} = 7$ TeV
 is measured to be $(98.3 \pm 0.2^{\rm stat} \pm 2.8^{\rm syst})$\,mb. 
 }

\begin{document}

\maketitle
\section{Introduction}
TOTEM is one of the special purpose experiments at CERN Large Hadron Collider (LHC), dedicated to make precision measurements of p+p scattering in the forward direction. TOTEM stands for TOTal and Elastic scattering cross-section Measurement.
The experiment is located at Interaction Point 5 at the LHC, shared with CMS,
which is one of the general purpose LHC experiments.

TOTEM consists of two inelastic telescope subsystems T1 and T2, 
located at approximately 10 and 14 meters from the collision point, and two Roman Pot (RP) detector stations, located approximately 147 and 220 m away from the 
collision point IP5, on its both sides in the LHC tunnel. 
A detailed description of TOTEM is published in ref.~\cite{TOTEM-0},
while the great physics potential of this specialized LHC experiment was
highlighted in ref.~\cite{TOTEM-00}. 
In this report, we review the results of the first two physics papers from TOTEM, 
based on refs.~\cite{TOTEM-01,TOTEM-02,TOTEM-03}. 

\section{Differential cross-section of elastic p+p scattering}
TOTEM is optimised for measuring elastic p+p scattering over a 
large four momentum transfer $|t|$-interval, ranging ultimately from
$10^{-3}$ to $10\,\rm GeV^{2}$, using specilazied LHC runs. 
The first TOTEM measurements of elastic pp scattering 
were reported in the $|t|$-range from 0.36 to 2.5\,GeV$^{2}$ in ref.~\cite{TOTEM-01}.
The data were taken using the standard 2010 LHC beam optics with 
$\beta^{*} = 3.5\,$m,  during a TOTEM dedicated run with four proton 
bunches of $7 \times 10^{10}$ p/bunch per beam with a total integrated luminosity of 6.1\,nb$^{-1}$.  This low-luminosity configuration allowed the Roman Pot detectors to approach the beams to a distance as small as 7 times the transverse beam size $\sigma_{\rm{beam}}$.

A reconstructed track in both projections in the near and in the far vertical RP 
unit is required on each side of the IP.
The two diagonals {\em top left of IP -- bottom right of IP} and 
{\em bottom left of IP -- top right of IP}, tagging possible elastic candidates, 
are used as almost independent experiments with slightly different optics corrections, 
yet constrained by the alignement of the RPs.  Collinearity cuts, acceptance, background and 
inefficiency corrections, resolution and bin migration effects, 
alignment and relevant LHC magnet dependent optics uncertainties are detailed in ref.~\cite{TOTEM-01}.  
The systematic uncertainty in $t$, the square of the four-momentum
transferred in the elastic scattering, was found to be dominated by optics and alignment.
The systematic uncertainties in d$\sigma$/d$t$ 
were dominated by the uncertainty on the efficiency correction (t-independent) and on the resolution unfolding, which depends on the $t$ measurement errors and hence on the uncertainty on the beam divergence.
The uncertainty of $|t|$  has been found to be $\delta |t| = 0.1$ GeV $\sqrt{|t|}$~\cite{TOTEM-01}. 

The time dependent instantaneous luminosity was taken from CMS 
measurements~\cite{int:lumi1,int:lumi2}, with uncertainty of $4\,\%$. 
The recorded luminosity is derived by integrating the luminosity, the trigger efficiency and the DAQ efficiency over all the different runs taken. 

\begin{figure}
\vspace{-0.4truecm}
\centering
\subfloat[The first TOTEM 
 data on differential cross-section of elastic p+p scattering at $\sqrt{s} = 7$ TeV,
 measured in the momentum transfer range of  0.36 $\le | t | \le $ 2.5 GeV$^{2}$. ]
 {\label{fig:1a}\includegraphics[width=0.45\textwidth]{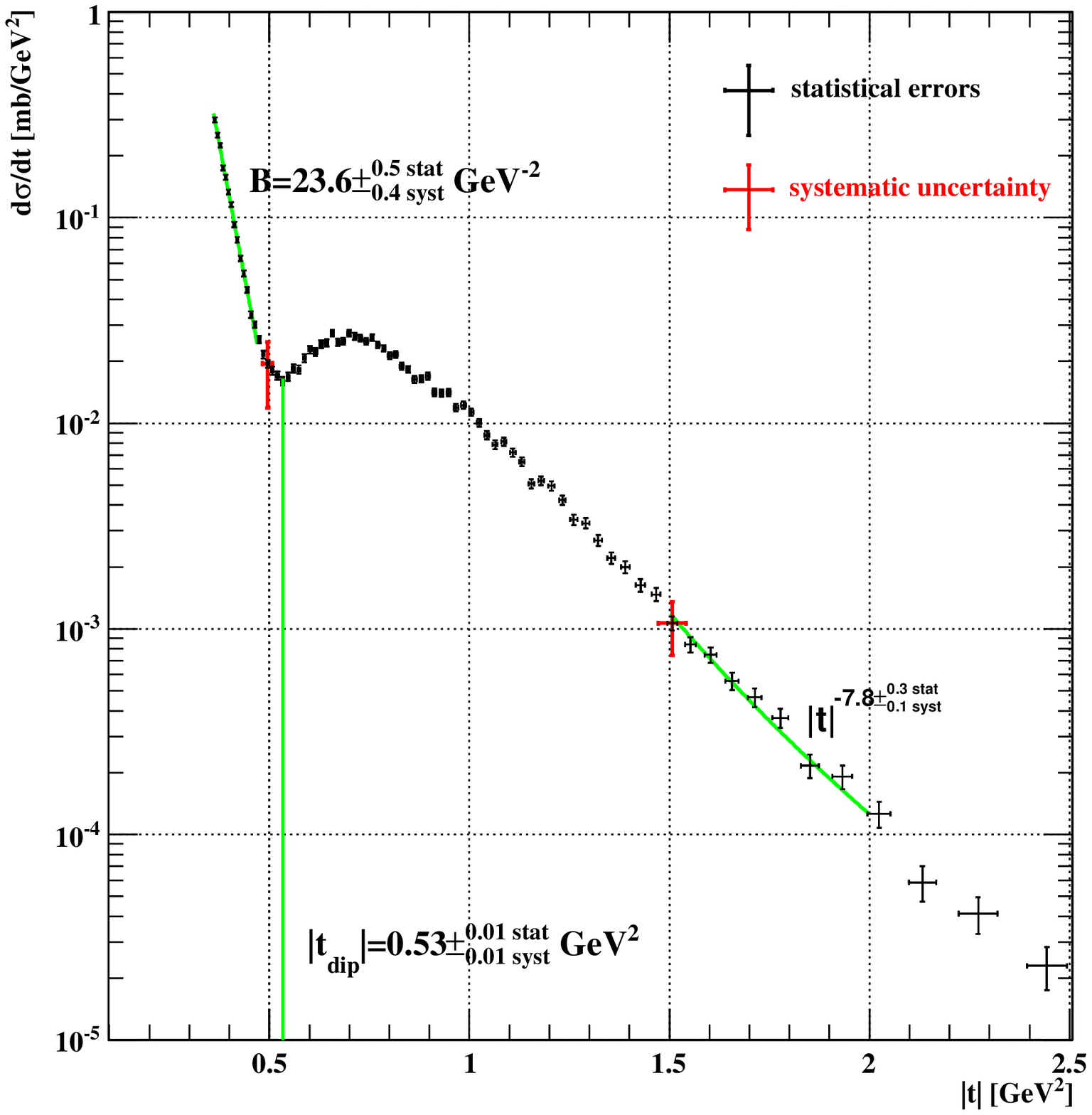}}
 \qquad
\subfloat[ When compared to predictions of different models, 
the TOTEM elastic scattering $d\sigma/dt$ data 
show a strong discriminative power~\cite{TOTEM-01}.]
{\label{fig:1b}\includegraphics[width=0.45\textwidth]{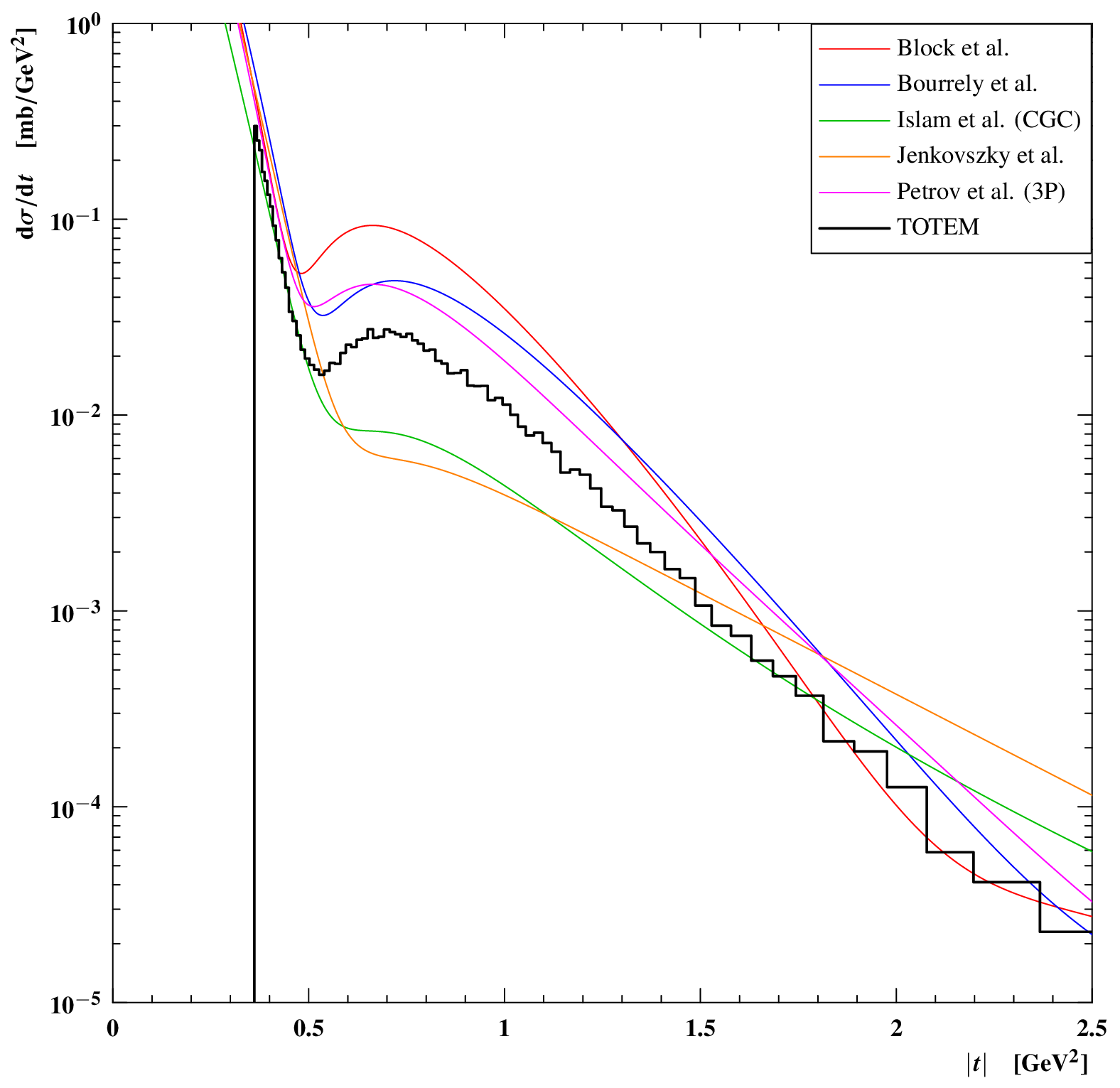}}
\caption{TOTEM elastic scattering $d\sigma/dt$ data measurements at $\sqrt{s} = 7$ TeV.}
\label{fig:1ab}
\end{figure}

The differential cross-section $\rm{d}\sigma/\rm{d}t$ 
for elastic p+p  
scattering at $\sqrt{s} = 7$ TeV is shown on the left panel of in Fig.~\ref{fig:1ab}.
Its comparison to model predictions is shown on the right panel of the same
Figure, indicating the precision and the high selectivity of the first TOTEM measurement.
For a more detailed discussion, see ref.~\cite{TOTEM-01}.

For further understanding of pp elastic scattering the $|t|$-range has to be considerably extended. 
The development of the approximately exponential behaviour at low $|t|$ is fundamental for 
the extrapolation to the optical point at $t$ = 0 and hence for the measurement of the 
elastic scattering and the total cross-section. 
The first TOTEM result on this topic is the subject of the next Section.
 
\section{Measurement of the total cross-section of p+p scattering}
TOTEM has also measured~\cite{TOTEM-02} the differential cross-section for elastic proton-proton 
scattering at $\sqrt{s} = 7$ TeV, analysing data from a short
run with a dedicated LHC optics of $\beta^{*} = 90 $ m.
This optics made it possible to measure  
the differential cross-section $\rm{d}\sigma/\rm{d}t$ 
of the elastic p+p scattering 
in the $|t|$ range of $(0.02 \div 0.33)\,\rm GeV^2$. 
In this range, a single exponential fit with a slope $B = (20.1 \pm 0.2^{\rm stat} \pm 0.3^{\rm syst})\,\rm GeV^{-2}$ describes the differential cross-section. 
(Note that in the $|t|$ range of 0.36 to 2.5 GeV$^2$ the slope is slightly different, 
$B = (23.6 \pm 0.5^{\rm stat} \pm 0.4^{\rm syst})\, \rm GeV^{-2}$, see Fig.~\ref{fig:1ab}.)
After the extrapolation of the data in the low $|t|$ range to $|t|=0$, a total elastic scattering cross-section of $(24.8 \pm 0.2^{\rm stat} \pm 1.2^{\rm syst})$\,mb was obtained.
Applying the optical theorem and using the luminosity measurement from CMS, 
and taking the COMPETE prediction as detailed in ref~\cite{TOTEM-02} for
the parameter $\rho = 0.14^{+0.01}_{-0.08}$ 
a total proton-proton cross-section of $(98.3 \pm 0.2^{\rm stat} \pm 2.8^{\rm syst})$\,mb was deduced. 
From the total and elastic pp cross-section measurements, an inelastic pp cross-section of $(73.5 \pm 0.6^{\rm stat} ~^{+1.8}_{-1.3}~^{{\rm syst}})$\,mb was also inferred.

\begin{figure}
\centerline{
\includegraphics[width=12cm,height=8cm]{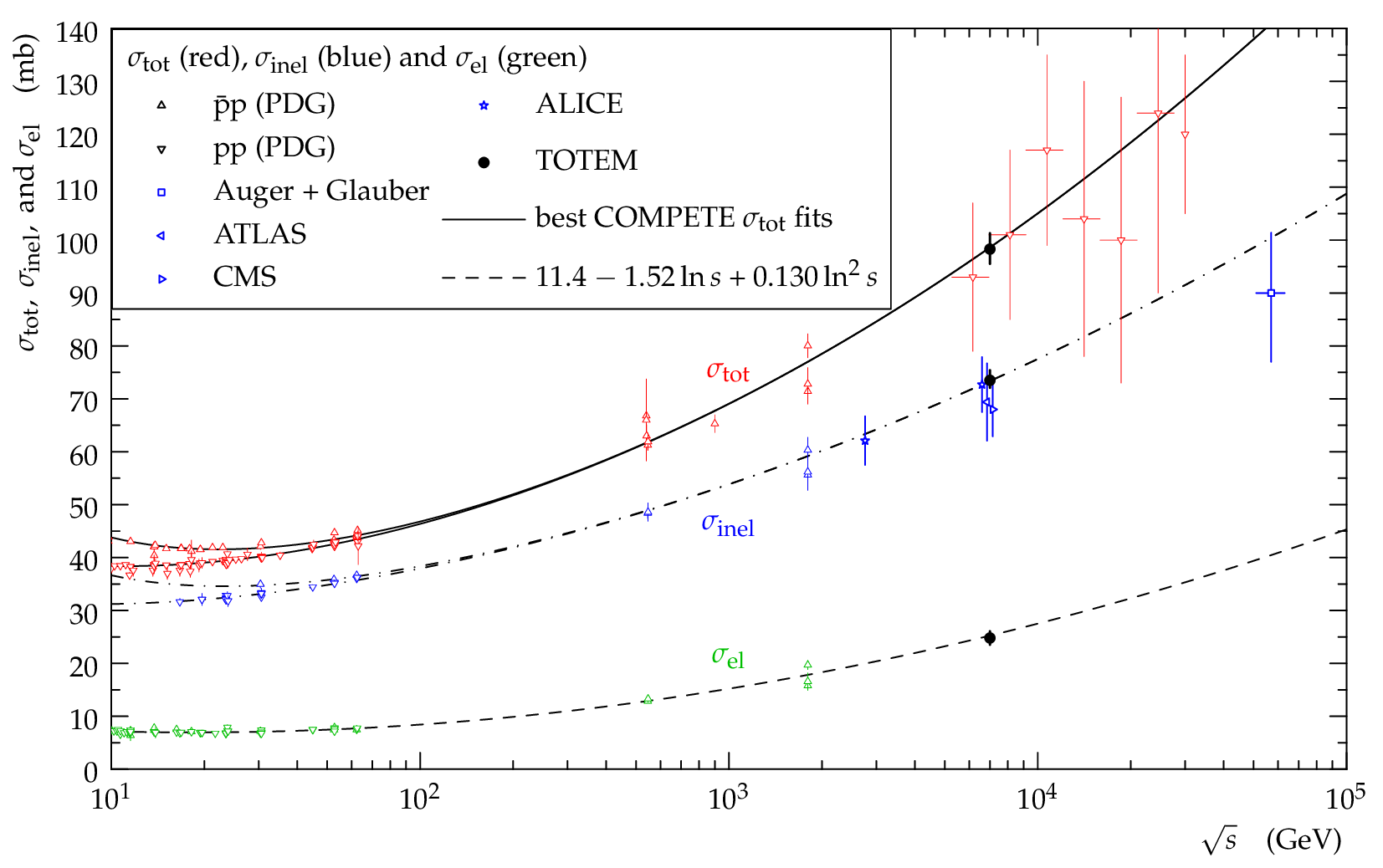}
}
\caption{
 The new TOTEM data demonstrate the continuation of the trends from earlier measurements,
 and indicate the high precision of the TOTEM experiment~\cite{TOTEM-02}.  }
 \label{figure:2}
\end{figure}

\section*{Acknowledgements}
T. Cs. would like to thank the Organizers of ISMD 2011 for providing a most enchanted and inspiring environment as well as for their patience,
and gratefully acknowledges support from KEK.
This work was supported by the TOTEM Institutions and also
by NSF (US), the Magnus Ehnrooth Foundation, Finland, the Waldemar von Frenckell Foundation, Finland, the Academy of Finland, and the Hungarian OTKA grants NK 73143, NK 101438 and the NKTH-OTKA grant 74458.

%

\end{document}